\documentclass[aps,prl,twocolumn,showpacs]{revtex4-1}
\usepackage{graphicx}
\usepackage{amsmath}
\usepackage{amssymb}
\usepackage{multirow}

\begin{document}
\title{Finite-Time-Finite-Size Scaling of the Kuramoto Oscillators}
\author{Mi Jin Lee}
\affiliation{Department of Physics, Sungkyunkwan University, Suwon 440-746, Korea}
\author{Su Do Yi}
\affiliation{Department of Physics, Sungkyunkwan University, Suwon 440-746, Korea}
\author{Beom Jun Kim}
\email[Corresponding author: ]{beomjun@skku.edu}
\affiliation{Department of Physics, Sungkyunkwan University, Suwon 440-746, Korea}

\begin{abstract}
We numerically investigate the short-time nonequilibrium temporal relaxation of
the globally-coupled Kuramoto oscillators, and  apply
the finite-time-finite-size scaling (FTFSS) method which contains
two scaling variables in contrast to the conventional
single-variable finite-size scaling.
The FTFSS method yields a smooth scaling surface, and the conventional finite-size
scaling curves can be viewed as proper cross sections of the surface.
The validity of our FTFSS method is confirmed by the critical
exponents in agreement with previous studies: Quenched disorder gives
the correlation exponent $\bar{\nu}=5/2$ and the dynamic exponent $\bar z = 2/5$
while thermal disorder leads to $\bar{\nu}=2$ and $\bar z = 1/2$, respectively.
We also report our results for the Kuramoto model
in the presence of both quenched and thermal disorder.
\end{abstract}
%05.45.Xt  : Synchronization; coupled oscillators
%05.70.Fh :	Phase transitions: general studies
%64.60.Ht  : Dynamic critical phenomena
%02.60.-x  : Numerical approximation and analysis

\maketitle

%\section{Introduction}
In the frame of statistical physics, it is important to find critical
exponents in a system aiming at understanding of its critical behaviors.
For a limited number of model systems, it might be possible to
analytically obtain the exponents via the mean-field analysis,
transfer-matrix calculation, the renormalization group approach, and
other analytic tools~\cite{golden}. In most of realistic model systems,
however, a rigorous analytic calculation of critical exponents is often a
formidable task, making numerical approaches unavoidable and essential.

A phase transition manifests itself as a singularity of the free energy,
which exists only in thermodynamic limit of the infinite system size. On the other
hand, one can only simulate finite systems in any computational approach,
which results in the so-called finite-size effects. The very finiteness of system
sizes can be utilized in the form of the finite-size scaling (FSS),
which has been successfully used to extract critical exponents~\cite{privman}.
In the conventional FSS, the system under study is asked to be in equilibrium,
and thus the functional form of FSS lacks any time dependence. As the system
size becomes larger, the time scale at criticality often diverges together
with the correlation length, which gives rise to the critical slowing
down. In order to circumvent this, the finite-size dynamical scaling
approach~\cite{zheng,soares} has been introduced. The basic assumption in this
approach of dynamical scaling is that one can extract critical behavior by
observing the temporal relaxation of the system in early times even
before reaching equilibrium.

\begin{figure}
\includegraphics[width=0.48\textwidth]{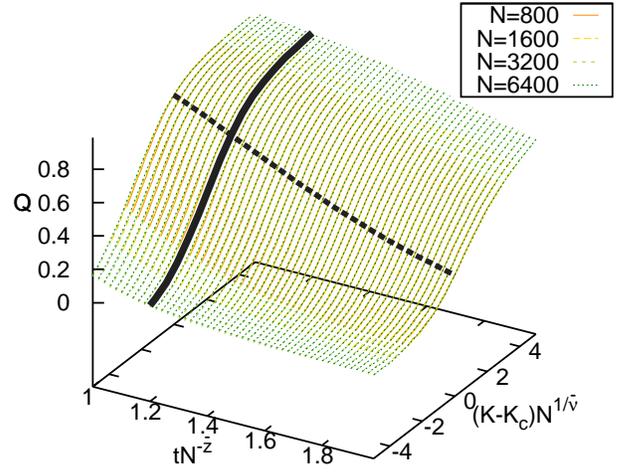}
\caption{(Color online) Finite-time-finite-size scaling (FTFSS):
$Q(t,N,K) = f\Bigl (t N^{\bar z},  (K-K_c)N^{1/\bar\nu}\Bigr )$
for the globally-coupled Kuramoto oscillators with quenched disorder
yields a smooth scaling surface with $\bar\nu = 5/2$, $\bar z = 2/5$, and $K_c = 1.595769$.
The scaling collapse is good enough to make difference of surfaces obtained from
different sizes $N=800, 1600, 3200$, and 6400 almost invisible.
The thick dashed and solid lines are two cross sections of the surface at $t N^{-\bar z} = 1.2$
and at $K=K_c$ displayed in Fig.~\ref{fig:quenched}(b) and (d), respectively. }
\label{fig:3dquenched}
\end{figure}

In general, one can assume that a time-varying thermodynamic quantity $Q$ is a function
of the time $t$, the linear size $L$, and the coupling strength $K$. As the critical
point $K = K_c$ is approached, the correlation length $\xi$ diverges following
the form $\xi \sim (K-K_c)^{-\nu}$ and so does the relaxation time scale $\tau
\sim \xi^{z}$. If we further assume that $Q$ is chosen in such a way
that its anomalous dimension is null~\cite{soares}, the scaling form
of $Q$ is written as
$Q(t,L,K) = f\Bigl (t L^{-z},  (K-K_c)L^{1/\nu}\Bigr )$.
In words, the first scaling variable $t L^{-z}$ describes the competition
of the two time scales, the finite observation time $t$ and the relaxation time $\tau$,
while the second scaling variable $(K-K_c)L^{1/\nu}$ is for the competition
of the two length scales, the finite system size $L$ and the correlation length
$\xi$.
It is straightforward to extend the scaling form for the globally-coupled system, which
reads
\begin{equation}
\label{eq:FTFSS}
Q(t,N,K) = f\Bigl (t N^{-\bar z},  (K-K_c)N^{1/\bar\nu}\Bigr ),
\end{equation}
where $N$ is the size of the system and $\bar \nu$ and $\bar z$ are
correlation and dynamic exponents defined for globally-coupled system.
Throughout the present study, we term the scaling in
Eq.~(\ref{eq:FTFSS}) as finite-time-finite-size scaling (FTFSS).
Precisely speaking, the phase transition is defined only in the limit of
infinite time $t \rightarrow \infty$ and infinite system size $N \rightarrow
\infty$, whereas any numerical calculation is limited by finiteness of $t$ and
$N$. Just like the conventional FSS aims to systematically utilizes
the finite-size effect by
using the scaling variable $L/\xi$, our FTFSS uses $t/\tau$ as well to
utilize the finite-time effect introduced by the finiteness of the observation
time $t$. Figure~\ref{fig:3dquenched} exhibits how the FTFSS can be used to produce a smooth
scaling surface for the globally-coupled Kuramoto model (see below for details).

{\it Model} --
Synchronization phenomena are ubiquitously observed in a variety of systems
such as the neuronal network and epilepsy in the brain~\cite{epilepsy,undesired},
circadian rhythm~\cite{circadian}, the collapse of the Millennium
Bridge~\cite{bridge}, power grids~\cite{powergrid0,powergrid}, and social
behavior of humans~\cite{heavymetal}.  Kuramoto model has been most popularly
used to describe such synchronization phenomena, and we study the globally-coupled
$N$ oscillators with both quenched intrinsic frequency
$\omega_i$ and thermal noise $\eta_i$ described by
\begin{equation}\label{eq:kura}
\frac{d\theta_i}{dt}=\omega_{i}-\frac{K}{N}\sum_{j=1}^{N}\sin(\theta_{i}-\theta_j)+\eta_{i}(t),
\end{equation}
where $K$ is the coupling strength and $\theta_i$ is the phase of the $i$-th
oscillator. We use the Gaussian distribution of the zero mean and the variance
$\sigma^2$ for the distribution function for $\omega_i$, and the thermal noise
satisfies $\langle \eta_i \rangle =0$ and $\langle \eta_i (t)\eta_j(t')\rangle
= 2T\delta_{ij}\delta(t-t')$ with $T$ being the effective temperature and
$\langle \cdots \rangle$ the ensemble average.  In the
zero-temperature limit of $T \rightarrow 0$, the system corresponds to the
conventional Kuramoto model for which the correlation exponent
$\bar{\nu}=5/2$~\cite{hong2007,mha} and the dynamic exponent $\bar z = 2/5$~\cite{mha} have
been found.  In the limit of $\sigma \rightarrow 0$, on the other hand, the
system behaves as the globally-coupled $XY$ model for which
$\bar{\nu}=2$~\cite{son2010,mha} and $\bar z = 1/2$~\cite{mha} are known.
This equation of motion~(\ref{eq:kura}) in a finite dimension can be
viewed as the time-dependent Ginzberg Landau (TDGL) dynamics of the superconducting
array with the dc current $\omega_i$ and the thermal noise current
$\eta_i$~\cite{kimTDGL}.

In order to carry out FTFSS, we use the initial condition $\theta_i(t=0) = 0$
for all oscillators and  measure the key quantity $Q(t)$ defined
by~\cite{kim3dxy,melwyn4dxy}
\begin{equation}
\label{eq:Q}
Q(t)
\equiv \left\langle
\textrm{sign}\left[\sum_{i=1}^{N}\cos \theta_{i}(t) \right] \right\rangle,
\end{equation}
which satisfies $Q(0)=1$ and $Q(t \rightarrow \infty) = 0$ at any parameter
values of $K$, $\sigma$, and $T$ due to the rotational $U(1)$ symmetry in
equilibrium.  The
advantage of using $Q$ lies on that it does not have the anomalous
dimension~\cite{soares}, which allows us to use the simple FTFSS form in
Eq.~(\ref{eq:FTFSS}).  We also emphasize that our FTFSS method uses how $Q(t)$
evolves in time before reaching equilibrium, and thus equilibration is not a
requirement to extract critical exponents~\cite{zheng}. However, it is to be
noted that before the ensemble average $\langle \cdots \rangle$, each
sample run returns the value either $+1$ or $-1$ at time $t$ since only the sign is taken in Eq.~(\ref{eq:Q}).
This means that in order to get a smooth continuous form for $Q(t)$ as a function
of time $t$, the ensemble average must be performed over sufficiently many samples.
We use the second-order algorithm to integrate equation of motion with the discrete
time step $\Delta t=0.01$.

\begin{figure}
\includegraphics[width=0.48\textwidth]{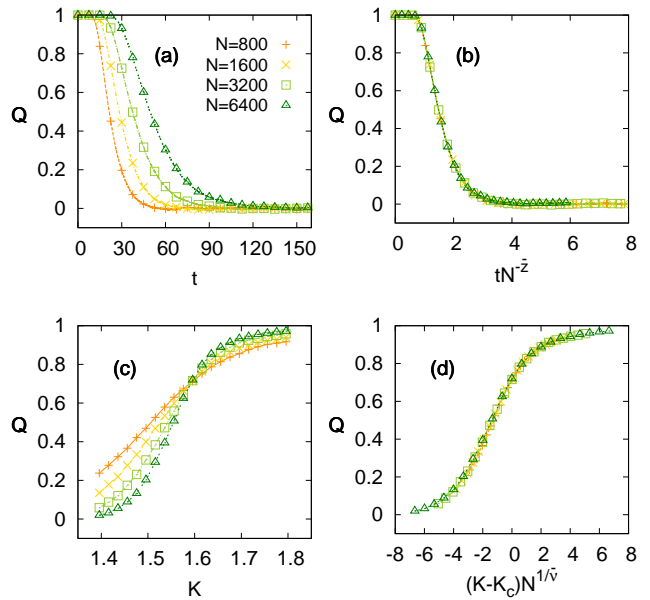}
\caption{(Color online)
The two-variable FTFSS form~(\ref{eq:FTFSS}) in Fig.~\ref{fig:3dquenched} at zero temperature is
cross sectioned to one variable scaling form (b) $Q = f\bigl (tN^{-\bar z},0\bigr )$ at $K=K_c$
and (d) $Q = f(tN^{-\bar z}=1.2, (K-K_c)N^{1/\bar\nu})$. The raw data [(a) and (c)] obtained for various
system sizes are scaled into smooth scaling curves [(b) and (d)] with the critical
exponents (b) $\bar z = 2/5$  and (d) $\bar \nu = 5/2$ with the critical coupling strength $K_c = 1.595769$.
}
\label{fig:quenched}
\end{figure}

{\it Results} --
We first report the scaling results obtained from 100000 samples for the usual Kuramoto model with only the quenched
randomness in the intrinsic frequency,
corresponding to the zero-temperature limit
of Eq.~(\ref{eq:kura}). The full FTFSS with the two scaling variables in Eq.~(\ref{eq:FTFSS})
is shown in Fig.~\ref{fig:3dquenched} with $\sigma = 1$. All surfaces obtained for various
sizes $N=800, 1600, 3200$, and 6400 collapse into a single smooth surface with the
critical exponents $\bar\nu = 5/2$, $\bar z = 2/5$ and the critical coupling strength
$K_c = \sqrt{8/\pi} = 1.595769$, as expected from previous studies~\cite{son2010,mha}.
The dynamic scaling~\cite{mha}
is easily obtained by fixing the second scaling variable in the FTFSS~(\ref{eq:FTFSS})
to null by putting $K = K_c$ as shown in Fig.~\ref{fig:quenched}(b). All the
curves in Fig.~\ref{fig:quenched}(a) are shown to collapse nicely into a single smooth
curve. Interesting application of the FTFSS is achieved by fixing the first scaling
variable $tN^{-\bar z}$ in Eq.~(\ref{eq:FTFSS}), to make the FTFSS form identical
to the conventional FSS form. As an example, we use
$\alpha \equiv tN^{-\bar z} = 1.2$ to plot Fig.~\ref{fig:quenched}(d). It is clearly
seen that by fixing the first scaling variable, one can successfully obtain the
critical exponent $\bar\nu = 5/2$ and $K_c = 1.595769$. In order to get such
a finite-size scaling as in Fig.~\ref{fig:quenched}(d), it is important to choose
the observation time $t$ systematically for the given system size $N$ to keep
the value $t N^{-\bar z}$ as constant.

\begin{figure}
\includegraphics[width=0.48\textwidth]{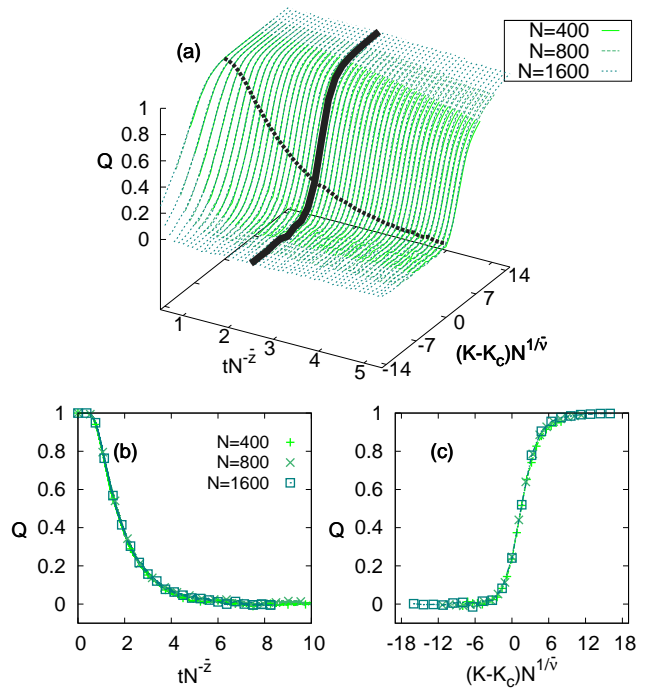}
\caption{(Color online) FTFSS for the globally-coupled
Kuramoto oscillators with thermal disorder only: $\bar\nu = 2$ and $\bar z = 1/2$
are obtained at $K_c = 2$ (in units of the temperature $T$).
The thick dashed and solid lines are two cross sections of the surface at $t N^{-\bar z} = 2.5$
and at $K=K_c$ displayed in (b) and (c), respectively.
}
\label{fig:3dthermal}
\end{figure}

In other extreme case with only thermal noise, corresponding to $\sigma = 0$ in
Eq.~(\ref{eq:kura}), the dynamics is effectively identical to the mean-field
version of the TDGL equation, for which it is known that $\bar\nu = \nu d_u=2$ and
$\bar z = z/d_u = 1/2$ beyond upper critical dimension $d_u = 4$~\cite{melwyn4dxy}.
In parallel to
Fig.~\ref{fig:quenched} where $K$ is in units of $\sigma$, we now measure $K$
in units of the temperature $T$.  From the well-known result that the critical
value of $T/K$ is 1/2 for the globally-coupled $XY$ model~\cite{globalXY},
we expect that $K_c = 2$ in units of the temperature $T$.
In the presence of thermal noise, integration of equation of motion takes much
longer time since generation of random number is needed at each time step.
We use $N=400$, 800, and 1600 and the ensemble averages are performed for 20000 samples.
As expected from known results of $K_c = 2$, $\bar\nu = 2$, and $\bar z = 1/2$, our
FTFSS gives us again a good quality of scaling surface as shown in Fig.~\ref{fig:3dthermal}(a).
As in Fig.~\ref{fig:quenched}, we also make cross sections of the scaling surface to construct
scaling collapses in Fig.~\ref{fig:3dthermal}(b) and (c), in which it is shown clearly
that the use of $\bar\nu = 2$ and $\bar z = 1/2$ with $K_c = 2$ yields scaling collapses
as expected.

\begin{figure}
\includegraphics[width=0.48\textwidth]{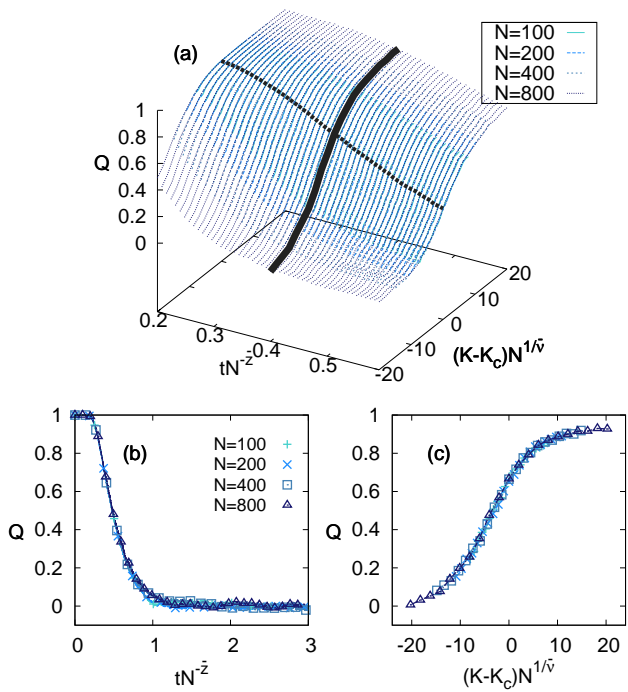}
\caption{(Color online) FTFSS for the globally-coupled
Kuramoto oscillators with both disorders: $\bar\nu = 2.22$ and $\bar z = 0.45$
are obtained at $K_c=5$ with $\sigma=2.28006$ (in units of the temperature $T$).
The thick dashed and solid lines are two cross sections of the surface at $t N^{-\bar z} = 0.4$
and at $K=K_c$ displayed in (b) and (c), respectively.
}
\label{fig:3dboth}
\end{figure}

%%%%%%%%%%%%%%%%%%%%%%%%%%%
\begin{table}[b]
\caption{The correlation exponent $\bar{\nu}$ is computed along the phase
boundary in the $(T, \sigma)$ plane (see Fig.~1 of~\cite{son2010}) via the
FTFSS of $Q$ and the FSS of $r$. The system sizes used here are 100, 200, 400,
and 800 with 10000 samples.  Estimation through the use of $Q$ gives us more
accurate values of $\bar\nu = 5/2$ and $\bar z = 2$ at $T=0$ and $T=0.5$,
respectively. We also include the value of $\bar z$ from the FTFSS of $Q$,
which satisfies $\bar z \approx 1/\bar\nu$.}
\begin{tabular}{c||c|c|c}
%    \multirow{2}*{$T/K$} & $\bar{\nu}$  & $\bar{\nu}$ & $\bar{z}$\\
%     & (from $r$) &(from $Q$) &  (from $Q$)\\
$T/K$ & $\bar{\nu}$ from $r$  & $\bar{\nu}$ from $Q$  & $\bar{z}$ from $Q$\\
    \hline
    $0$ & $2.35$ & $2.5$ & $0.4$ \\
    $0.1$ & $2.23$ & $2.38$ & $0.42$ \\
    $0.2$ & $2.11$ & $2.22$ & $0.45$ \\
    $0.3$ & $2.07$ & $2.08$ & $0.48$ \\
    $0.4$ & $2.03$ & $2.04$ & $0.49$ \\
    $0.5$ & $1.98$ & $2$ & $0.5$ \\
\end{tabular}
\label{tab:summary}
\end{table}

Motivated by the success of our FTFSS method for the Kuramoto model in the presence of either
purely quenched disorder (Figs.~\ref{fig:3dquenched} and \ref{fig:quenched}) or
purely thermal disorder (Fig.~\ref{fig:3dthermal}), we next study the Kuramoto systems
with both types of disorder ($\sigma \neq 0$ and $T \neq 0$). 
In Fig.~\ref{fig:3dboth}, we show the FTFSS of $Q$
at $\sigma/T = 2.28006$ gives us a good quality of scaling collapse with 
$\bar\nu = 2.22$, $\bar z = 0.45$, and $K_c/T = 5$ 
(the values for $\sigma/T$ and $K_c/T$ are taken on the phase boundary in~\cite{son2010}).
It has been suggested that the upper critical dimension $d_u = 5$
for the Kuramoto model with quenched randomness~\cite{hong2005}, and $d_u = 4$ has been
agreed for the globally-coupled $XY$ model~\cite{globalXY}, which explains the values
obtained above ($\bar\nu = 5/2, \bar z = 2/5$ for quenched,  and $\bar\nu = 2, \bar z = 1/2$ for thermal
disorder, respectively)
on equal footing via $\bar\nu = \nu d_u$ and $\bar z = z/d_u$ with $\nu = 1/2$ and $z = 2$.
The phase diagram in the two-dimensional parameter space of $(T, \sigma)$ with both in units of $K$
has been obtained
in~\cite{son2010}. We apply the FTFSS for our key quantity $Q$ in the same way as above
to find the critical exponents along the phase boundary,
and compare with the results obtained from the standard finite-size scaling of the Kuramoto order
parameter $r$ defined by
\begin{equation}
r \equiv \left\langle\left|  \frac{1}{N}\sum_i e^{i\theta}\right|  \right\rangle ,
\end{equation}
where $\langle \cdots \rangle$ is for both the sample average and the temporal average after equilibration.
The standard FSS form
$r = N^{-\beta/\bar{\nu}}g\bigl( (T-T_c) N^{1/\bar\nu}\bigr)$
is then used with the critical exponent $\beta=1/2$~\cite{son2010} for the order parameter.
In Table~\ref{tab:summary} we compare $\bar\nu$ obtained from the FTFSS of $Q$
and from the FSS of $r$, along the phase boundary presented
in Fig.~1 of~\cite{son2010}. In the zero-temperature limit, the correct value
$\bar\nu = 5/2$ is well obtained from $Q$,
but the value is rather inaccurate if $r$ is used instead.
In other limit of $\sigma = 0$ for which the phase boundary crosses $T$ axis at 1/2,
the value $\bar\nu \approx 2$ from $Q$ is more accurate than the one from $r$ although the difference is
not as significant as in the zero-temperature case. Another advantage of the FTFSS of $Q$
is that it allows us to obtain other exponent $\bar z$, which is also listed in Table~\ref{tab:summary}.
It is noteworthy that along the full phase boundary, $\bar z = 1/\bar\nu$ is found regardless
of the value of $T/K$, which appears to suggest that 
$\bar z \bar \nu =  (z/d_u)(\nu d_u) = z \nu = 2 \cdot (1/2) = 1$
remains constant along the phase boundary.
Standard universality argument suggests that as soon as thermal disorder is added
to the Kuramoto system $\bar\nu$ is expected to change abruptly from 5/2 to 2.
Although our results in Table~\ref{tab:summary} do not show such an abrupt change
of $\bar\nu$,
we believe that this can be an finite-size artifact which might disappear if much bigger system
sizes are used. In spite of the advantage of using the FTFSS of $Q$ that only
initial stage of nonequilibrium short-time relaxation is enough to detect universality
class, we point out that the calculation of $Q$ requires average over a larger
number of samples. In contrast, the calculation of $r$, for large systems in particular,
does not require extensive sample average, thanks to the self-averaging property.

{\it Conclusion and Summary} --
We have applied the finite-time-finite-size scaling (FTFSS)
for the globally-coupled Kuramoto model with quenched and
thermal disorders.
The key quantity $Q(t)$ has been defined as
the ensemble average of the sign of the real part of the Kuramoto order
parameter, and measured as a function of time $t$.
We have found that the FTFSS with the two scaling variables
yields a well-defined scaling surface, cross sections of which in two different
directions lead to the standard finite-size scaling as well as the dynamic scaling.
Correlation critical exponent $\bar\nu$ and the dynamic critical exponent $\bar z$
have been obtained through the use of FTFSS applied for the quantity $Q(t)$,
confirming the results from previous studies: $\bar\nu = 5/2, \bar z = 2/5$ for
purely quenched randomness and $\bar\nu = 2, \bar z = 1/2$ for purely
thermal noise. Our FTFSS of $Q(t)$ is based on the early stage of nonequilibrium
relaxation and thus makes it possible to avoid the critical slowing down near the criticality.
%Recently, Ref.~\cite{mha} has also discussed the dynamic scaling in the
%globally-coupled Kuramtmo model.
%The authors' basic notions are very similar to ours other than that they use
%the order parameter for the scaling and deterministically assign the intrinsic
%frequencies by the random sampling~\cite{son2010} in order to get rid of the
%quenched disorder.  The obtained values of the exponents $\bar{\nu}$ and
%$\bar{z}$ in the quenched disorder case by~\cite{mha}are exactly matched to the
%values in this paper. Just like our study, their research gives a support to
%the validity and utility of the dynamic scaling in short-time regime in the
%globally-coupled Kuramoto model.

%In the generalized case $T$ seems to contribute to the continuous change of
%universality class, more specifically the upper critical dimension, of the
%system. We do not make sure yet whether indeed $T$ works in the way against
%existing opinion on the effect of $T$ yet. Despite of this vagueness, we
%confirmed the SFSS is powerfully effective method for the whole case.
%{\bf AAAAAAAAAAAAA}

This work was supported by the National Research Foundation of Korea (NRF) grant
funded by the Korea government (MEST) (No. 2011-0015731).


\begin{thebibliography}{}
\bibitem{golden} N. Goldenfeld, {\it Lectures on Phase Transitions and the Renormalization Group}, (Addison-Wesley, New York, 1992)
\bibitem{privman} D. Privman, {\it Finite Size Scaling and Numerical Simulation of Statistical Systems}, (World Scientific, Singapore, 1990).
\bibitem{zheng} Z. B. Li, L. Sch\"{u}lke, and B. Zheng, Phys. Rev. Lett. {\bf 74}, 3396 (1995).
\bibitem{soares} M. S. Soares, J. K. L. da Silva, F. C. S. Barreto, Phys. Rev. B {\bf 55}, 1021 (1997).
\bibitem{mha} C. Choi, M. Ha, and B. Kahng, arXiv:1307.2408v1 (2013).
\bibitem{epilepsy} R. S. Fisher, W. v. E. Boas, W. Blume, C. Elger, P. Genton, P. Lee, and J. Engel, Epilepsia {\bf 46}, 470 (2005).
\bibitem{undesired}  V. H. P. Louzada,  N. A. M. Ara\'{u}jo, J. S. Andrade Jr., and H. J. Herrmann, Sci. Rep. {\bf 2}, 658 (2012).
\bibitem{circadian} E. Kolmos and S. J. Davis, Current Biology, {\bf 17}(18), 808 (2007).
\bibitem{bridge}  S. H. Strogatz, D. M. Abrams, A. McRobie, B. Eckhardt, and E. Ott, Nature, {\bf 438}, 43. (2005).
\bibitem{powergrid0} G. Filatrella, A. H. Nielsen, and N. F. Pedersen, Eur. Phys. J. B {\bf 61}, 485 (2008).
\bibitem{powergrid} M. Rohden, A. Sorge, M. Timme, and D. Witthaut, Phys. Rev. Lett. {\bf 109}, 064101 (2012).
\bibitem{heavymetal} J. L. Silverberg, M. Bierbaum, J. P. Sethna, and I. Cohen, Phys. Rev. Lett. {\bf 110}, 228701 (2013).
\bibitem{hong2007} H. Hong, H. Chat\'{e}, H. Park, and L.-H Tang, Phys. Rev. Lett. {\bf 99}, 184101 (2007).
\bibitem{son2010} S.-W. Son and H. Hong, Phys. Rev. E {\bf 81}, 061125 (2010).
\bibitem{kimTDGL} B. J. Kim, P. Minnhagen, and P. Olsson, Phys. Rev. B {\bf 59}, 11506 (1999).
\bibitem{kim3dxy} B. J. Kim, L. M. Jensen, and P. Minnhagen, Physica B {\bf 284}, 413 (2000).
\bibitem{melwyn4dxy} L. M. Jensen, B. J. Kim, and P. Minnhagen, Physica B {\bf 284}, 455 (2000).
\bibitem{globalXY} S. K. Baek and B. J. Kim, Phys. Rev. E {\bf 86}, 011132 (2012); B. J. Kim, H. Hong, P. Holme, G. S. Jeon, P. Minnhagen, and M. Y. Choi, Phys. Rev. E {\bf 64}, 056135 (2001); M. Antoni and S. Ruﬀo, Phys. Rev. B {\bf 52}, 2361 (1995).
\bibitem{hong2005} H. Hong, H. Park, and M. Y. Choi, Phys. Rev. E {\bf 72}, 036217 (2005).



\end{thebibliography}
\end{document}

\title{Finite-Time and Finite-Size Scaling of Relaxation Behavior: Supplemental Material}
\author{Mi Jin Lee}
\affiliation{Department of Physics, Sungkyunkwan University, Suwon 440-746, Korea}
\author{Su Do Yi}
\affiliation{Department of Physics, Sungkyunkwan University, Suwon 440-746, Korea}
\author{Beom Jun Kim}
%\email[Corresponding author: ]{beomjun@skku.edu}
\affiliation{Department of Physics, Sungkyunkwan University, Suwon 440-746, Korea}

\maketitle

\section{Globally-Coupled Kuramoto Model with Quenched and Thermal Disorder}

\begin{figure}
\includegraphics[width=0.68\textwidth]{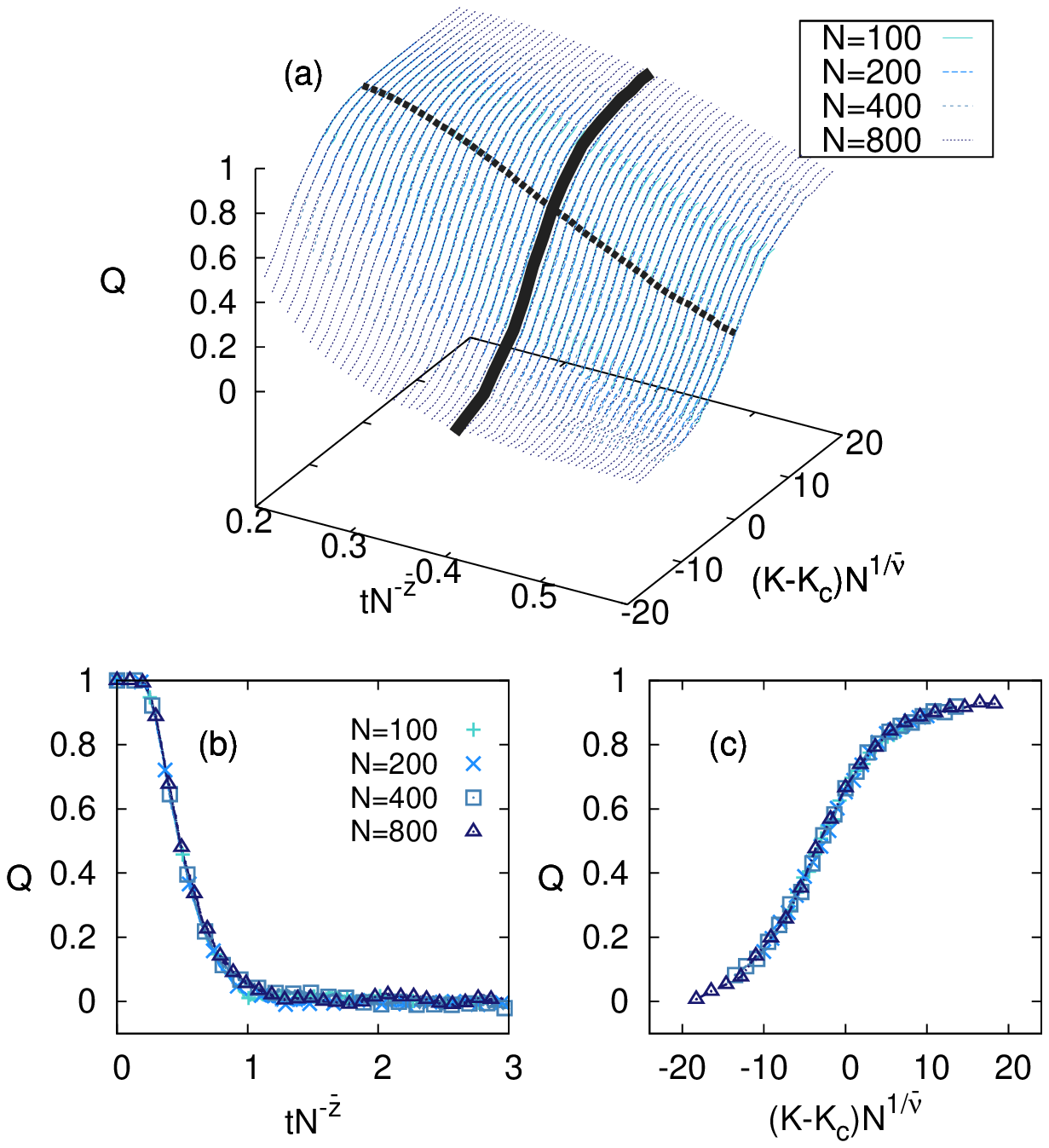}
\caption{FTFSS for the globally-coupled
Kuramoto oscillators with both quenched and thermal disorder: $\bar\nu = 2.3$ and $\bar z = 0.45$
are obtained at $K_c=5$ with $\sigma=2.28006$ (both in units of the temperature $T$).
The thick dashed and solid lines are two cross sections of the surface at
$K=K_c$ and at $t N^{-\bar z} = 0.4$, which are  displayed in (b) and (c), respectively.
}
\label{fig:3dboth}
\end{figure}

\begin{figure}
\includegraphics[width=0.48\textwidth]{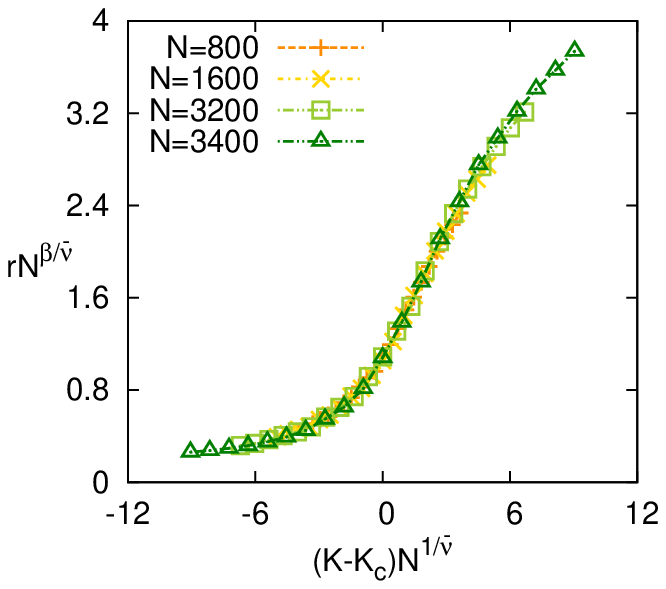}
\caption{The standard FSS of the order parameter $r$
for the globally-coupled Kuramoto oscillators at zero temperature. With $\beta = 1/2$ and $K_c = 1.595769$,
a scaling collapse is obtained at $\bar \nu = 2.3$.
}
\label{fig:rscale}
\end{figure}

Motivated by the success of our FTFSS method for the Kuramoto model in the
presence of either purely quenched disorder (see Figs.~1 and 2 in the main
paper) or purely thermal disorder (see Fig.~3 in the main paper), we hereby
study the Kuramoto systems with both types of disorder [$\sigma \neq 0$ and $T
\neq 0$ in Eq.~(2) in the main paper].  In Fig.~\ref{fig:3dboth}, we show the
FTFSS of $Q$ at $\sigma/T = 2.28006$ gives us a good quality of scaling
collapse with $\bar\nu = 2.3$, $\bar z = 0.45$, and $K_c/T = 5$ (the values
for $\sigma/T$ and $K_c/T$ are taken from the phase boundary in~\cite{son2010}).
It has been suggested that the upper critical dimension $d_u = 5$ for the
Kuramoto model with quenched randomness~\cite{hong2005}, and $d_u = 4$ has been
agreed for the globally-coupled $XY$ model~\cite{globalXY}, which explains the
values obtained in the main paper ($\bar\nu = 5/2, \bar z = 2/5$ for quenched,  and
$\bar\nu = 2, \bar z = 1/2$ for thermal disorder, respectively) on equal
footing via $\bar\nu = \nu d_u$ and $\bar z = z/d_u$ with $\nu = 1/2$ and $z =
2$.
We apply the
FTFSS to our key quantity $Q$ in the same way as in the main paper to find the critical
exponents along the phase boundary, and compare with the results obtained from
the standard finite-size scaling (FSS) of the Kuramoto order parameter $r$ defined by
\begin{equation}
r \equiv \left\langle\left|  \frac{1}{N}\sum_i e^{i\theta_i}\right|  \right\rangle ,
\end{equation}
where $\langle \cdots \rangle$ is for both the sample average and the temporal average in the steady state.
The standard FSS form
$r = N^{-\beta/\bar{\nu}}g\bigl( (T-T_c) N^{1/\bar\nu}\bigr)$
is then used with the critical exponent $\beta=1/2$~\cite{son2010} for the order parameter.
In Fig.~\ref{fig:rscale}, as an example, we display the standard FSS of $r$ measured in equilibrium at zero
temperature, corresponding to Figs.~1 and 2 in the main paper. A good quality of the scaling collapse is achieved
at $\bar \nu \approx 2.3$ at $\beta = 1/2$.
%It is to be noted that the cross section of the FTFSS surface
%of $Q$ gives us more accurate value of $\bar \nu = 2.5$.

\begin{table}[b]
\caption{The correlation exponent $\bar{\nu}$ is computed along the phase
boundary in the $(T, \sigma)$ plane via the
FTFSS of $Q$ and the FSS of $r$. The system sizes used here are 100, 200, 400,
and 800 and the averages over 10000 samples are performed.
We also include the values of $\bar z$ from the FTFSS of $Q$,
which satisfy $\bar z \approx 1/\bar\nu$. We expect that if much bigger
system sizes are used, $\bar \nu$ and $\bar z$ converge to 2.0 and 0.5,
respectively, except at the zero temperature limit for which $\bar \nu = 2.5$ and $\bar z = 0.4$
are expected. The numbers in parentheses are
errors in the last digits.
}
\begin{tabular}{c||c|c|c}
$T/K$ & $\bar{\nu}$ from $r$  & $\bar{\nu}$ from $Q$  & $\bar{z}$ from $Q$\\
    \hline
    $0$ &   $2.3(1)$ & $2.5(1)$   & $0.40(1)$ \\
    $0.1$ & $2.2(1)$ & $2.4(1)$   & $0.42(1)$ \\
    $0.2$ & $2.1(1)$ & $2.3(1)$   & $0.45(1)$ \\
    $0.3$ & $2.1(1)$ & $2.1(1)$   & $0.49(1)$ \\
    $0.4$ & $2.0(1)$ & $2.1(1)$   & $0.50(1)$ \\
    $0.5$ & $2.0(1)$ & $2.0(1)$   & $0.51(1)$ \\
\end{tabular}
\label{tab:summary}
\end{table}

In Table~\ref{tab:summary} we compare $\bar\nu$ obtained from the FTFSS of $Q$
and from the FSS of $r$, along the phase boundary presented
in Fig.~1 of~\cite{son2010}.
When the randomness in the system is purely thermal ($\sigma=0$)
for which the phase boundary crosses $T/K$ axis at 1/2,
the value $\bar\nu \approx 2$ is observed both from $Q$ and $r$.
In the zero-temperature limit ($T/K = 0$), on the other hand,
the correct value $\bar\nu = 5/2$ is well obtained from $Q$,
but the value is rather inaccurate if $r$ is used instead.
Another advantage of using the FTFSS of $Q$
is that it allows us to obtain other exponent $\bar z$, which is also listed in Table~\ref{tab:summary}.
It is noteworthy that along the full phase boundary, $\bar z \approx 1/\bar\nu$ is found regardless
of the value of $T/K$, which appears to suggest that
$\bar z \bar \nu =  (z/d_u)(\nu d_u) = z \nu = 2 \cdot (1/2) = 1$
remains constant along the phase boundary.
In spite of the advantage of using the FTFSS of $Q$ that only
initial stage of nonequilibrium short-time relaxation is enough to detect universality
class, we point out that the calculation of $Q$ requires average over a larger
number of samples. In contrast, the calculation of $r$, for large systems in particular,
does not require extensive sample average, thanks to the self-averaging property.

\begin{figure}
\includegraphics[width=1.0\textwidth]{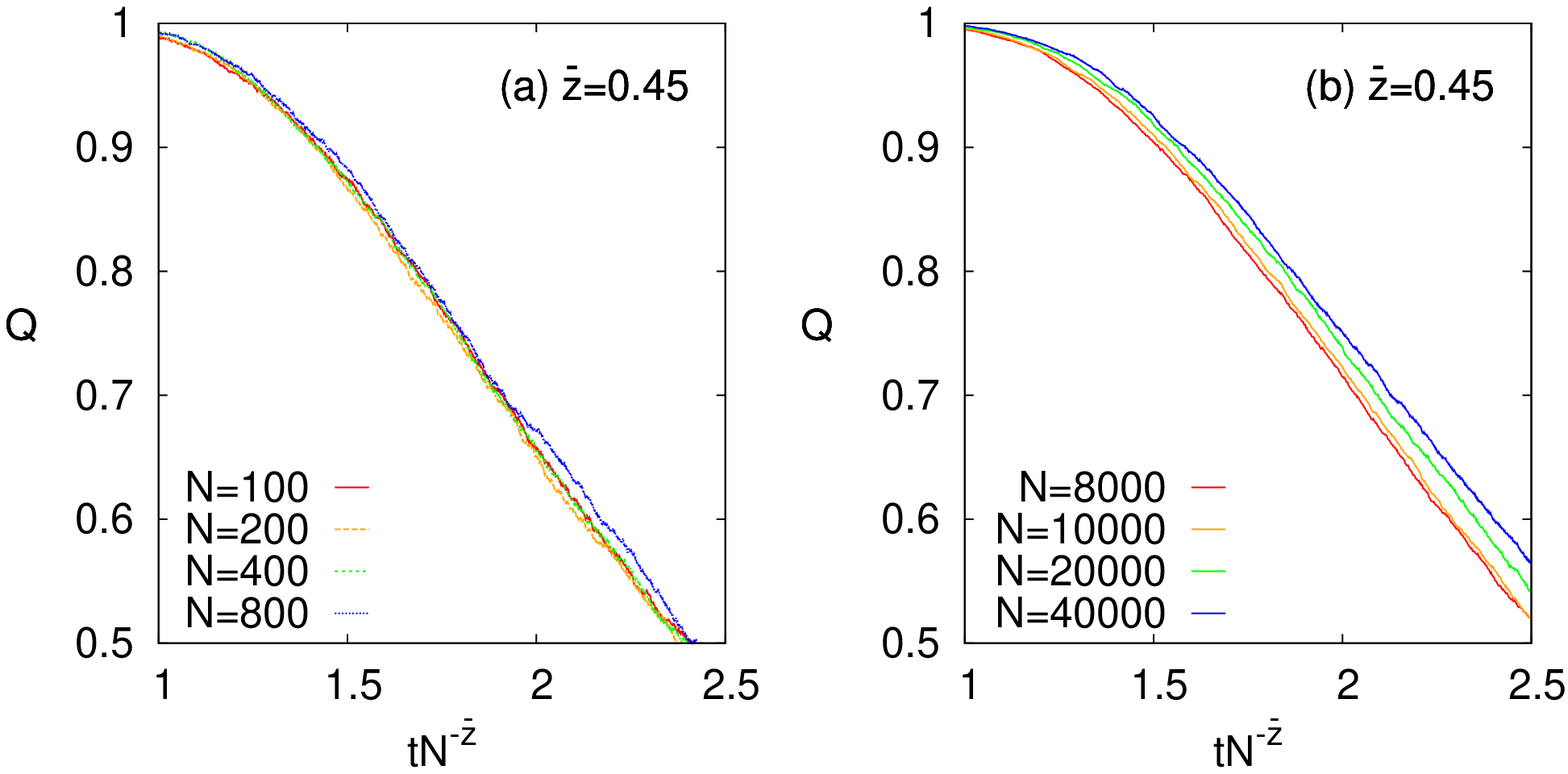}
\caption{The cross section scaling at $K=K_c$ of the FTFSS
surface of the globally-coupled Kuramoto oscillators for $T/K = 0.2$. (a) For
relatively small system sizes $N=100, 200, 400$, and 800, $\bar z=0.45$ gives
us the best scaling collapse.  (b) For larger system sizes $N=8000$, $10000$,
$20000$, and $40000$, the quality of the scaling for the same value $\bar z =
0.45$ is shown to become worse, whereas the use of the larger value $\bar z =
0.47$ yields much better scaling collapse as shown in (c).
}\label{fig:zlarge}
\end{figure}

Standard universality class argument suggests that as soon as thermal disorder is
added to the Kuramoto system, the correlation exponent is expected to change from $\bar \nu =5/2$  to $\bar \nu = 2$,
and the dynamic exponent $\bar z = 2/5$ to $\bar z = 1/2$.  Although our results in Table~\ref{tab:summary} do not show such abrupt changes of $\bar\nu$ and $\bar z$, we believe that this can be a finite-size artifact
which might be remedied if much bigger system sizes are used.  In order to check
this, we fix $T/K=0.2$ and compute $Q(t)$ at the critical point $K_c$ for larger sizes $N=8000$, 10000, 20000, and 40000
to compute the dynamic critical exponent $\bar{z}$. As displayed in Fig.~\ref{fig:zlarge},
the bigger system sizes yield the larger value of $\bar{z}$, which appears to suggest that
the expected value $\bar z = 1/2$ can be obtained if much bigger system sizes are used for
the FTFSS analysis.

\section{Kuramoto oscillators with quenched randomness in the Watts-Strogatz small-world network}

\begin{figure}
\includegraphics[width=0.7\textwidth]{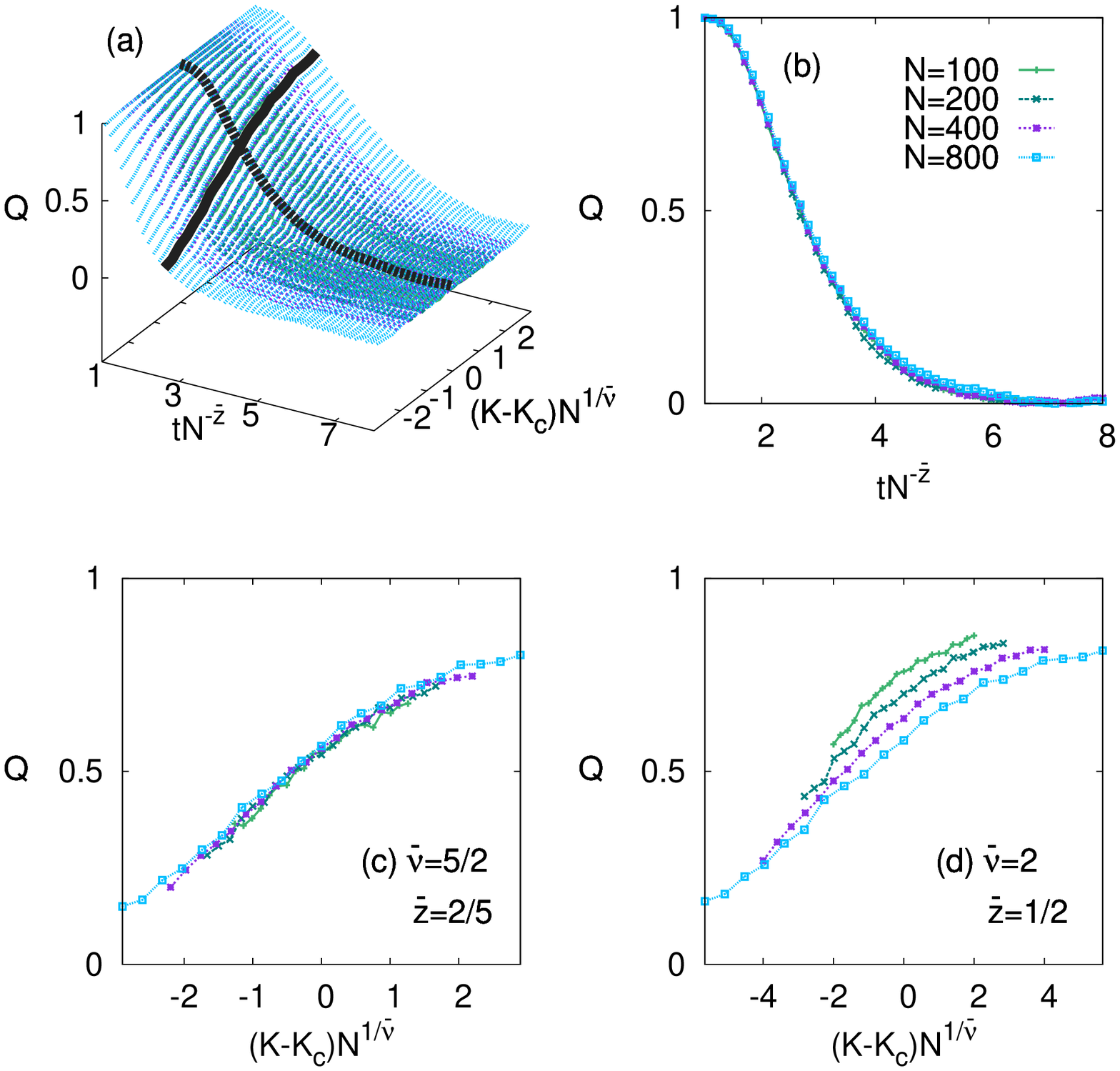}
\caption{(a) The FTFSS of $Q$ for the Kuramoto oscillators in
the WS network at the rewiring probability $p=0.5$. The two cross sections for
$K=K_c$ and $tN^{-\bar z} = 2.54$, respectively, denoted as two thick lines in
(a), exhibit scaling collapses in (b) and (c) with $\bar{z}=2/5$ and
$\bar{\nu}=5/2$. (d) The scaling collapse at $t N^{-\bar z} = 1.28$
with $\bar z = 1/2$ and $\bar \nu = 2$, corresponding to $d_u = 4$
with $z = 2$ and $\nu = 1/2$,  becomes much worse than in (c).
}
\label{fig:ws}
\end{figure}

As one of examples of complex network structures, we use the Watts-Strogatz(WS)
small-world network as an underlying interaction structure of the Kuramoto oscillators.
The WS network has drawn much interest by playing the role of a bridge between
the regular and random networks, which is controlled by the rewiring
procedure~\cite{wsnetwork}. For a rewiring probability $p=0$, the WS network
has a regular ring lattice structure, while it becomes a fully random network
for $p=1$.  For the intermediate value of $p$ ($0<p<1$) the structure of the WS
network is well described as the small-world network, characterized by the
short path length and the large clustering coefficient~\cite{wsnetwork}.  It is
well known that for any nonzero value of $p$, the universality class of the WS
network becomes identical to the globally-coupled system~\cite{smallsynch}.  We
generate the WS networks at $p=0.2, 0.5$, and $0.8$ with the average degree
fixed as $\langle k \rangle = 6$ and integrate equations of motion [Eq.~(4) in
the main text] for the Kuramoto oscillators.  Our FTFSS method is then applied to
$Q(t)$, as shown in Fig.~\ref{fig:ws}(a).  In Fig.~\ref{fig:ws}(b) and (c), we
exhibit the cross sections [denoted as thick full and dotted lines in
Fig.~\ref{fig:ws}(a)].  As expected, the WS network and the globally-coupled
structure are shown to have the same critical exponents (results for
$p=0.2$ and $p=0.8$ are not included in SM): $\bar{\nu}=2.5$ and
$\bar{z}=0.4$~\cite{hong2005}.  In Fig.~\ref{fig:ws}(d), we show the scaling along
the cross section for $tN^{-\bar z} = 1.28$ similarly to Fig.~\ref{fig:ws}(c),
but with $\bar z = 1/2$ and $\bar \nu = 2$. Different from \cite{smallsynch}, 
we conclude that the use of $\bar z = 2/5$ and $\bar \nu = 5/2$
as in Fig.~\ref{fig:ws}(c) yields better scaling collapse, confirming
$d_u = 5$.

\section{Monte-Carlo Dynamics of the Globally-Coupled $q$-state clock model}

To confirm the validity of FTFSS, we carry out similar procedure for the
globally-coupled $q$-state clock model, in which each planar spin
$\mathbf{S}_i=(\cos \theta_i, \sin \theta_i)$ has a
discrete direction $\theta_i =2\pi n_i/q$ with $n_i=0, ..., q-1$ and the Hamiltonian
is written as
\begin{equation}
\mathcal{H}=-J \sum_{i,j=1}^N\mathbf{S}_i\cdot\mathbf{S}_j
\end{equation}
with the coupling strength $J$. In the limit of
$q\rightarrow \infty$, the model corresponds to the $XY$ model, and when $q=2$
it is exactly identical to the Ising model.
In the $q$-state clock model, the competition between thermal fluctuation
and ferromagnetic interaction  produces the phase transition from disordered to ordered
phase~\cite{notq3,qglobal}. It is well known that for $q \neq 3$ the system exhibits  the
continuous phase transition of the mean-field nature. When $q=3$, on the other hand,
there is a discontinuous phase transition~\cite{q3}.

\begin{figure}
\includegraphics[width=0.8\textwidth]{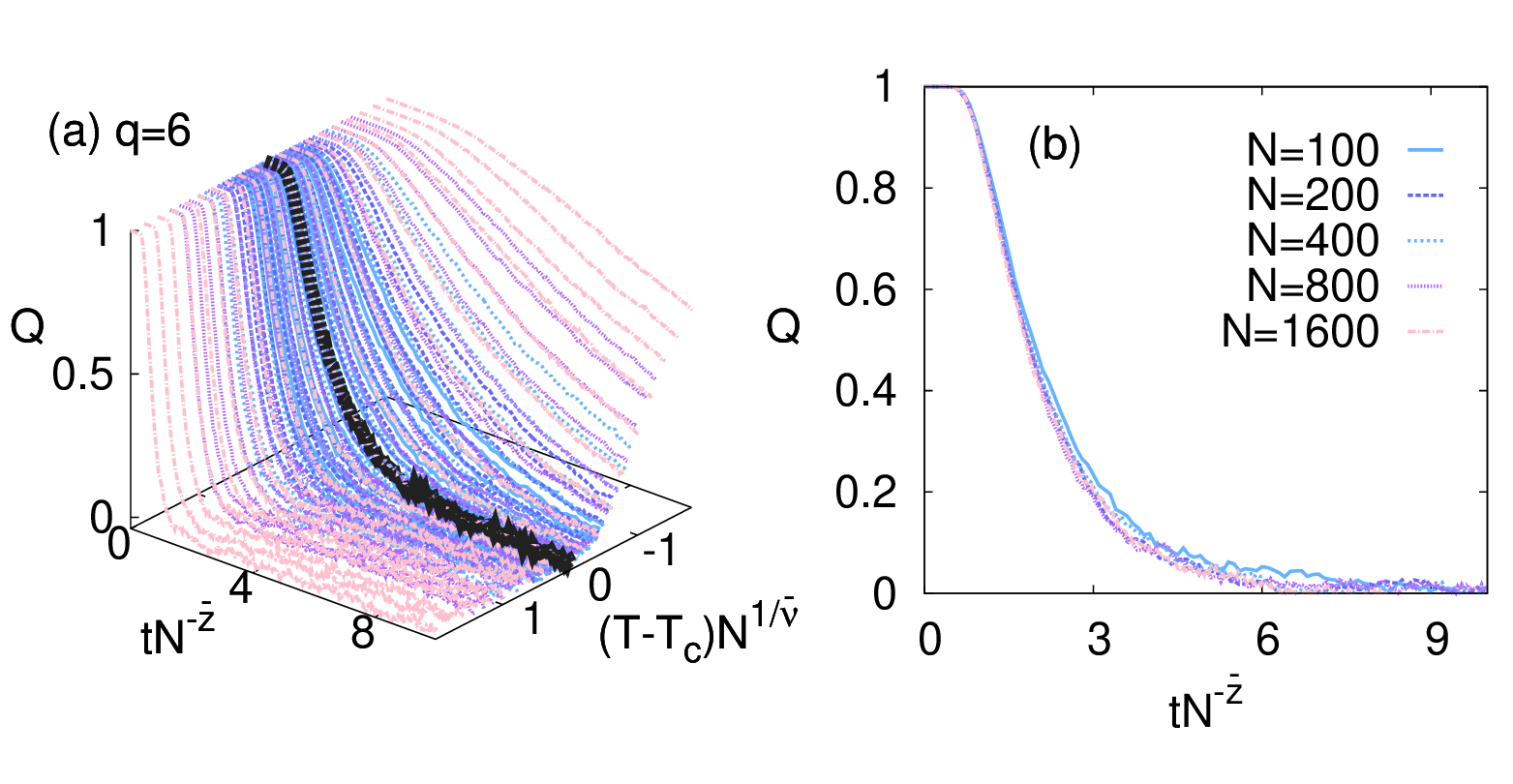}
\caption{(a) The FTFSS surface for the globally-coupled $q$-state clock model
at $q=6$. With $T_c = 1/2$ in units of $J$~\cite{qglobal} and
the well-known mean-field exponents $\bar{z}=1/2$ and $\bar{\nu}=2$, a good quality of
scaling surface is obtained. (b) Cross section of the scaling surface at $T=T_c$ [denoted as thick full
line in (a)].}
\label{fig:qstate}
\end{figure}

In contrast to the Kuramoto model, dynamics of the $q$-state clock model
is not solely specified by the Hamiltonian and the Monte-Carlo (MC) dynamics
depends on the choice of the spin update rule. We use the
the heat-bath algorithm and measure the time $t$ in units of
one MC step, which corresponds to $N$ MC tries of the spin update.
We present our FTFSS for the globally-coupled $q$-state clock model
with $q$=2, 4, 6, and 16 in Fig.~5 of the main paper.
In Fig.~\ref{fig:qstate}(b) in the present SM, we display the scaling
for $q=6$ along the cross section denoted in Fig.~\ref{fig:qstate}(a).  The choice of
the well-known mean-field exponents $\bar{z}=1/2$ and $\bar{\nu}=2$ gives us  again
a good quality of the scaling collapse, validating our FTFSS method also for
MC dynamics.